\documentclass[superscriptaddress,floatfix,preprintnumbers, nofootinbib,hyperref]{revtex4-2} 
\pdfoutput=1
\usepackage[title]{appendix}
\usepackage[colorlinks=true,breaklinks=true]{hyperref}
\usepackage[normalem]{ulem}
\usepackage[utf8]{inputenc}
\hypersetup{allcolors=[rgb]{0.0 0.0 0.6},linkcolor=[rgb]{0.75 0.05 0.05}}
\usepackage{amsmath,amssymb}
\usepackage{epsfig}  
\usepackage{graphicx}   
\usepackage{slashed}       
\usepackage{url}
\usepackage[dvipsnames]{xcolor}
\usepackage{color}
\usepackage{multirow}
\usepackage{comment}
\usepackage{amssymb}
\usepackage[version=3]{mhchem}
\usepackage{orcidlink}
\usepackage{longtable}

\DeclareUnicodeCharacter{2212}{-}

\allowdisplaybreaks

\bibpunct{[}{]}{,}{n}{}{,}
\definecolor{darkspringgreen}{rgb}{0.09, 0.45, 0.27}

\begin{document}

\vspace*{-1.5 truecm}
\begin{figure}[htbp]
\centering
    \includegraphics[width=4cm]{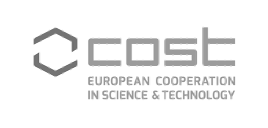}
    \hspace{1.truecm}
    \includegraphics[width=4cm]{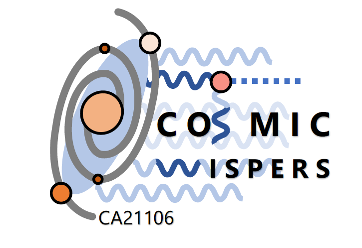}
    \hspace{1.5truecm}
    \includegraphics[width=3cm]{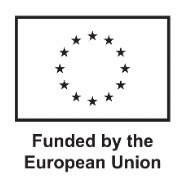}
\end{figure}

\vspace{1.4cm}

\title{Exploring the Dark Universe:\\ A European Strategy for Axions and other WISPs Discovery\\} 

\author{Deniz Aybas}
\affiliation{Department of Physics, Bilkent University, Ankara, Turkey 06800}

\author{María Benito}
\affiliation{Tartu Observatory, University of Tartu, Observatooriumi 1, Toravere 61602, Estonia}

\author{Francesca Calore}
\affiliation{LAPTh, CNRS, F-74000 Annecy, France}
\affiliation{LAPP, CNRS, F-74000 Annecy, France}

\author{Michele Cicoli}
\affiliation{Dipartimento di Fisica e Astronomia, Universit\`a di Bologna, via Irnerio 46, 40126 Bologna, Italy}
\affiliation{INFN, Sezione di Bologna, viale Berti Pichat 6/2, 40127 Bologna, Italy}

\author{Arturo de Giorgi}
\affiliation{Institute for Particle Physics Phenomenology, Durham University, South Road, DH1 3LE, Durham, UK}

\author{Amelia Drew}
\affiliation{The Abdus Salam International Centre for Theoretical Physics (ICTP), Strada Costiera 11, 34151 Trieste, Italy}
\affiliation{Institute for Fundamental Physics of the Universe (IFPU), Via Beirut 2, 34151 Trieste, Italy}

\author{Silvia Gasparotto}
\affiliation{Institut de F\'isica d’Altes Energies (IFAE), The Barcelona Institute of Science and Technology, Campus UAB, 08193 Bellaterra (Barcelona), Spain}
\affiliation{Grup de F\'{i}sica Te\`{o}rica, Departament de F\'{i}sica, Universitat Aut\`{o}noma de Barcelona, 08193 Bellaterra (Barcelona), Spain}

\author{Claudio Gatti}
\affiliation{INFN, National Institute for Nuclear Physics, I-00044, Frascati, Italy}

\author{Maurizio Giannotti}
\affiliation{Centro de Astropart{\'i}culas y F{\'i}sica de Altas Energ{\'i}as. University of Zaragoza. Zaragoza, \textit{50009}, Spain}
\affiliation{Department of Chemistry and Physics. Barry University. Miami Shores, \textit{33161}, Florida, USA}

\author{Marco Gorghetto}
\affiliation{Deutsches Elektronen-Synchrotron DESY, Notkestr. 85, 22607 Hamburg, Germany}

\author{Mathieu Kaltschmidt}
\affiliation{CAPA \& Departamento de Física Teórica, Universidad de Zaragoza, C. Pedro Cerbuna 12, 50009 Zaragoza, Spain}

\author{Marin Karuza}
\affiliation{Physics Department, University of Rijeka, 51000 Rijeka, Croatia}

\author{Alessandro Lella}
\affiliation{Dipartimento Interateneo di Fisica  ``Michelangelo Merlin'', Via Amendola 173, 70126 Bari, Italy}
\affiliation{INFN, Sezione di Bari, Via Orabona 4, 70126 Bari, Italy}

\author{Giuseppe  Lucente}
\affiliation{SLAC National Accelerator Laboratory, 2575 Sand Hill Rd, Menlo Park, CA 94025}

\author{Alessandro Mirizzi}
\email{alessandro.mirizzi@ba.infn.it}
\affiliation{Dipartimento Interateneo di Fisica  ``Michelangelo Merlin'', Via Amendola 173, 70126 Bari, Italy}
\affiliation{INFN, Sezione di Bari, Via Orabona 4, 70126 Bari, Italy}

\author{Mario Reig}
\affiliation{Rudolf Peierls Centre for Theoretical Physics, University of Oxford, Parks Road, Oxford OX1 3PU, United Kingdom}

\author{Nicole Righi}
\affiliation{Physics Department, King’s College London,
Strand, London, WC2R 2LS, U.K.
}
\author{Ophir M. Ruimi}
\affiliation{Racah Institute of Physics, Hebrew University of Jerusalem, 9190401 Jerusalem, Israel}
\affiliation{Johannes Gutenberg-Universität Mainz, 55128 Mainz, Germany}
\affiliation{Helmholtz Institute Mainz, 55099 Mainz, Germany}
\affiliation{GSI Helmholtzzentrum für Schwerionenforschung GmbH, 64291 Darmstadt, Germany}

\author{Elisa Todarello}
\affiliation{School of Physics and Astronomy, University of Nottingham, University Park, NG7 2RD, Nottingham, United Kingdom}

\author{Edoardo Vitagliano}
\affiliation{Dipartimento di Fisica e Astronomia, Universit\`a degli Studi di Padova, Via Marzolo 8, 35131 Padova, Italy}
\affiliation{INFN, Sezione di Padova, Via Marzolo 8, 35131 Padova, Italy}

\begin{abstract}
Axions and other very weakly interacting slim (with $m <$ 1 GeV) particles (WISPs) are a common feature of several extensions of the Standard Model of Particle Physics. The search of WISPs was already recommended in the last update of the European strategy on particle physics (ESPP). After that, the physics case for  WISPs has gained additional momentum. Indeed, WISPs may provide a new paradigm to explain the nature of dark matter and puzzling astrophysical and particle physics observations. This document briefly summarizes current searches for WISPs and the perspectives in this research field for the next decade, ranging from their theoretical underpinning, over their indirect observational consequences in astrophysics, to their search in laboratory experiments. It is stressed that in Europe a rich, diverse, and low-cost experimental program is already underway with the potential for one or more game-changing discoveries. In this context, it is also reported the role of the EU funded COST Action \emph{``Cosmic WISPers in the Dark Universe: Theory, astrophysics, and experiments''} (CA21106, \url{https://www.cost.eu/actions/CA21106}) in coordinating and supporting WISPs searches in Europe, shaping a roadmap to track the strategy to guarantee a European leadership in this field of research. This document has been submitted in March 2025 as an input to the update process of the ESPP.
\end{abstract}
\smallskip

\maketitle

\newpage 

\section{Introduction}

The Standard Model (SM) of Particle Physics is an extremely successful theory, explaining all interactions between known elementary particles as observed in the laboratory to a very high accuracy. However, there is enormous evidence about the existence of physics Beyond the SM (BSM). Notably, the SM fails to offer satisfactory explanations for the values of its numerous parameters, lacks a mechanism to unify gravity with quantum mechanics, and does not account for the origin of dark energy and dark matter (DM). In this context, for a long time weakly interacting massive particles (WIMPs) have constituted the dominant paradigm proposed to address some of these questions, in particular providing suitable candidates for DM particles. At this regard, the opportunity to investigate the electroweak scale using the Large Hadron Collider (LHC) has naturally led  DM research in recent years to focus on WIMP models. 

As a result, alternative DM candidates have received comparatively less attention from the experimental community. However, till now both  LHC searches and (in)direct detection experiments have not provided any clear evidence for the existence of WIMPs. The absence of unambiguous signals of new physics in these experiments motivates the recommendation expressed in the 2020 update of the European Strategy on Particle Physics (ESPP) to broaden the experimental efforts, exploring off-the-beaten tracks in comparison with traditional BSM physics explored at the LHC. In this context, bosonic very weakly interacting slim particles (WISPs) are emerging as  motivated DM alternatives to WIMPs~\cite{Jaeckel:2010ni}.
Ultraviolet (UV) SM completions can account for different types of WISPs,  that can be identified according to their spin:
\begin{itemize}
\item[-] \emph{Spin 0:} Pseudo Nambu-Goldstone bosons may arise from the breaking of well-motivated BSM symmetries at a scale much larger than the electroweak one. The prime example is the pseudo-scalar axion, introduced to solve the charge conjugation and parity (CP) problem in Quantum Chromodynamics (QCD), through the Peccei-Quinn mechanism. In a certain mass range, the axion can also provide a viable DM candidate. Nevertheless, beside the ``QCD axion'', there could be other pseudo-scalar particles with properties very similar to the axion. These axion-like particles (ALPs) emerge naturally in many theoretically appealing UV completions of the SM, as in the low-energy 4-dimensional limit of string theory compactified on Calabi-Yau threefolds \cite{Cicoli:2012sz}. Note that these UV complete frameworks include additional scalar particles like dilatons and string moduli which tend to be very light and to couple to ordinary matter with gravitational strength. These particles are suitable candidates to address several existing challenges in fundamental physics like DM, but also inflation and dark energy \cite{Cicoli:2023opf}. 

\item[-] \emph{Spin 1:} Another rather generic feature of UV completions of the SM is the presence of hidden sector $U(1)$ gauge theories under which SM particles are uncharged. If massive, these so-called ``hidden'' or ``dark'' photons, can mix kinetically via loops with the SM hypercharge, and so with the ordinary photons \cite{Abel:2008ai}. Ordinary matter then develops a tiny charge under the hidden photon and, in turn, hidden sector fermions, if present in the model, can acquire a small charge under the ordinary electromagnetic $U(1)_{\rm em}$.

\item[-] \emph{Spin 2:} ``Hidden'' or ``dark'' massive gravitons occur in bigravity which is the unique ghost-free bimetric extension of General Relativity. Bigravity may emerge as a low-energy manifestation of double copies in field theoretic UV completions and in string theory. They also appear as a necessary byproduct of extra-dimensional theories as Kaluza-Klein modes of the metric. Focusing instead on the ordinary graviton, fundamental physics can lead to a plethora of mechanisms to produce gravitational waves in early universe cosmology.
\end{itemize}

All these WISPs offer a convincing physics case in connection to the puzzle of DM, and provide a variety of opportunities for experimental and observational searches. Recent years have seen a renewed interest towards WISPs with impressive developments on both the theoretical and the experimental side. A rich,
diverse, and low-budget experimental program is already underway and has the potential for one or more game-changing discoveries~\cite{Irastorza:2018dyq}.
European groups are currently leading this research activity, with the largest particle physics laboratories (like CERN, DESY, LNF)  directly involved in the experimental searches of WISPs. This vivid experimental activity is supported by stimulating theoretical investigations, where new ideas and unexplored approaches to study WISPs have been proposed~\cite{Choi:2024ome}.
Moreover, in the last years there has been an intense investigation on the cosmological role of WISPs~\cite{Arias:2012az,OHare:2024nmr}.
Furthermore, the impact on different astrophysical observables has been analyzed ~\cite{Caputo:2024oqc,Carenza:2024ehj}.

In this document we provide an overview of the current searches for WISPs and the perspectives in this research field for the next decade, ranging from theory and model building (Sec.~\ref{sec:WP1}), to cosmology and DM (Sec.~\ref{sec:WP2}), astrophysics (Sec.~\ref{sec:WP3}) and direct searches in laboratory (Sec.~\ref{sec:WP4}).

This contribution is based on work carried on by the EU funded COST Action
\emph{``COSMIC WISPers in the Dark Universe: Theory, astrophysics and experiments''} (CA21106; \url{https://www.cost.eu/actions/CA21106}). 
The aim of the Action is to organize the scientific foundation for the next generation of WISPs experiments and searches in Europe. Apart from promoting cooperation and coordination, a crucial responsibility of the Action is to strengthen a specific European WISP strategy in the context of {ESPP}, launching a
roadmap process. Currently there are about 450 researchers from (extra)-European countries that are affiliated with the Action, to carry on WISP studies on a systematic and coherent basis in the intersecting and interdisciplinary area of particle physics, astrophysics and cosmology.   

\section{WISPS THEORY AND MODEL BUILDING}
\label{sec:WP1}

From the point of view of string phenomenology, the intensity frontier should definitely be a top priority of the ESPP given that each naive string compactification yields naturally new physics in the form of very light and feebly coupled particles. These can be spin-0 modes like string moduli and axion-like particles \cite{Cicoli:2012sz}, spin-1 states like hidden photons which kinetically mix with the Standard Model hypercharge \cite{Abel:2008ai}, and hidden sectors more in general which could very well include light hidden fermions. Moreover, there are several ways in string cosmology to produce spin-2 particles in the form of gravitational waves (GWs). A non-exhaustive illustrative list of examples involves \cite{Cicoli:2023opf}: primordial tensor modes during inflation, particle production during inflation, secondary GWs associated to primordial black hole (PBH) production, GWs at non-perturbative preheating, GWs from a thermal bath of highly excited fundamental superstrings, GWs from a cosmological network of cosmic strings, GWs from oscillons, and GWs from phase transitions. 

Focusing in particular on axions, we are at a crucial moment in the history of axion physics: geometrical and computational tools have advanced to a stage where we can systematically determine the properties of the axions (their number, mass spectrum and couplings) which arise from compactifications of string theory, and at the same time astrophysical observations and terrestrial experiments have reached a sophistication and a scale such that an actual axion detection might conceivably occur relatively soon. Thus, it is of fundamental importance to keep aiming for experiments focused on the detection of the QCD axion, axion-like particles and hidden photons, as well as GW observations, focusing in particular on high frequencies. 

Let us stress that it is also crucial to have a complementary approach focused on the search for new physics in the energy frontier with the new generation of particle accelerators at CERN, as the FCC. Despite recent disillusion, low-energy supersymmetry is not ruled out yet, especially since in the MSSM the tree-level Higgs mass is below the $W$ mass, and so loop corrections need to be rather large. Hence it is not so strange to expect that the masses of coloured superparticles (squarks and gluini) are of order a few TeV. In addition, from the point of view of string theory, the most natural way to generate soft terms is via gravity mediation. However, in these scenarios the mass of the superpartners is of order of the mass of the moduli which needs to be $m_{\rm mod}\gtrsim\mathcal{O}(30)$ TeV to avoid cosmological problems (to make the moduli decay before Big Bang Nucleosynthesis (BBN) without ruining its successful predictions). Therefore, this more stringy view point strengthens the expectation to have superpartner a bit above the TeV scale.

\section{WISPS Cosmology and Dark Matter}
\label{sec:WP2}

\begin{figure}[t!]
    \begin{center}
    \includegraphics[width=0.9\columnwidth]{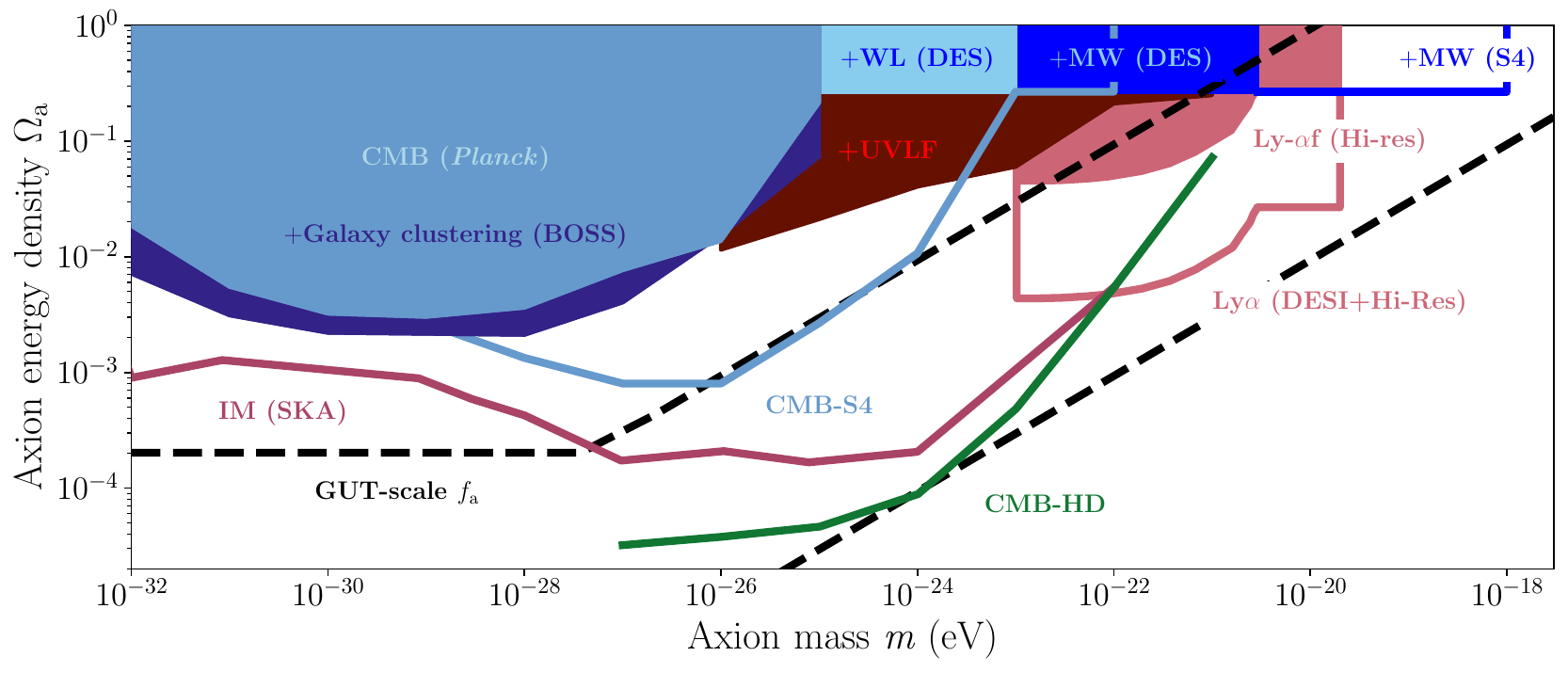}
        \caption{A summary of current cosmological bounds (shaded regions) on ultralight ALP mass \(m\) and relic density \(\Omega_\mathrm{a}\) with projected sensitivities (solid lines). Credits~\href{https://keirkwame.github.io/DM_limits/}{Keir K. Rogers}. See text for more details.}
       \label{fig:limits}
    \end{center}
\end{figure}


Axions and WISPs are also among the leading candidates for dark matter and dark energy, which together constitute roughly 95\% of the universe's energy budget. While these particles arise automatically as dark matter relics and can act as vacuum energy, a precise understanding of their production mechanisms, cosmological evolution, and observational signatures is still lacking to this day. Advancing our knowledge on this is essential for rigorously testing WISP models against existing and upcoming cosmological data, and for possibly opening up new WISP discovery pathways.

Pinning down precisely WISP production mechanisms in the early universe remains a key priority. For QCD axions and ALPs, these 
fall into two main categories: pre-inflationary and post-inflationary mechanisms, each with distinct phenomenological implications and detection prospects. In the pre-inflationary case, axions are produced non-thermally via the misalignment mechanism. A major challenge in this scenario is calculating the axion potential around the QCD crossover, which introduces uncertainties in the axion mass range for which this particle constitutes the observed dark matter density. New, precise determinations of the QCD topological susceptibility at finite temperature via lattice QCD simulations are necessary to improve on this uncertainty.

Conversely, in the post-inflationary scenario axions are produced from the decay a network of topological defects, axion strings and domain walls. For the QCD axion, this scenario is, in principle, completely predictive in terms of the axion mass. However, computational limitations and discrepancies between different methods of extrapolation to cosmological scales introduce significant uncertainties in the uniquely-predicted axion mass. This parameter is a critical input to resonance searches, and its precise determination could accelerate axion discovery by years. Additionally, this scenario leads to the formation of inhomogeneous axion distributions at sub-galactic scales, resulting in unique substructures like axion miniclusters and stars.  Understanding the formation and evolution of these objects is crucial for refining predictions of the dark matter distribution within galactic halos, directly affecting its density in the Solar System (key for direct detection) and influencing a range of cosmological and astrophysical observations.

Dark photons and dark gravitons are also compelling dark matter candidates. While several production mechanisms have been proposed, much of their viable mass range and SM couplings remain unexplored in term of minimal production scenarios. 


Axions and WISPs can also be produced in the early universe through interactions with the SM thermal bath, e.g. with photons or gluons. Depending their (still uncertain) thermalization rates, these particles may contribute to the DM as a \textit{hot} or \textit{warm} component, constrained e.g. by Cosmic Microwave Background (CMB) measurements for the effective number of relativistic degrees of freedom ($\Delta N_{\mathrm{eff}}$). Pinning down the precise WISP mass and couplings that are allowed by these cosmological constraints is crucial to identifying the viable parameter space for both direct and indirect searches, while ensuring accurate predictions to compare with upcoming observational data.


The cosmological evolution of WISPs leaves detectable imprints in the CMB and the growth of density perturbations, as illustrated in Fig.~\ref{fig:limits}. Many existing constraints rely on the suppression of the matter power spectrum on scales below the Jeans scale. As shown in Fig.~\ref{fig:limits}, a combination of different observables already imposes stringent constraints on the cosmological abundance of ALPs, particularly \textit{disfavoring} ALPs with masses $m \lesssim 10^{-20}$ eV as a dominant dark-matter component. 
Investigating higher masses with cosmological probes is challenging, as it requires a detailed understanding of the ultralight DM-induced Jeans suppression of cosmological perturbations, as well as its wave effects across various astrophysical and cosmological contexts. Progress in this area will require dedicated field-theory and N-body simulations that combine numerical approaches and approximate analytical methods, alongside machine learning techniques to facilitate comparisons between simulations and observational data. 

At yet smaller masses $m\lesssim H_0\simeq 10^{-33}$ eV, WISPs are promising candidates for quintessence dark energy. This is particularly relevant in light of recent observations suggesting evolving dark energy, based on the combined analysis of the CMB, Type Ia Supernovae (SNe Ia) and the recent Baryon Acoustic Oscillations (BAO) data~\cite{DESI:2024mwx,DESI:2025zgx}. WISPs have also been proposed as potential solutions to several outstanding tensions in cosmology, including the $H_0$
and $\sigma_8$ discrepancies, and naturally provide an explanation for the observed birefringence in the CMB polarization.

As outlined, cosmology offers a powerful framework for WISP searches, not only because these particles are strong candidates for dark matter and dark energy, but also due to their rich phenomenology, which enables key predictions -- such as the axion mass -- and yields observable signatures in cosmological data. Tackling the open challenges requires a multidisciplinary approach that combines theoretical advancements with the development of new computational tools and advanced data analysis techniques.


\section{WISPs Astrophysics}
\label{sec:WP3}

\begin{figure}[t!]
  \begin{center}
    \includegraphics[totalheight=7cm]{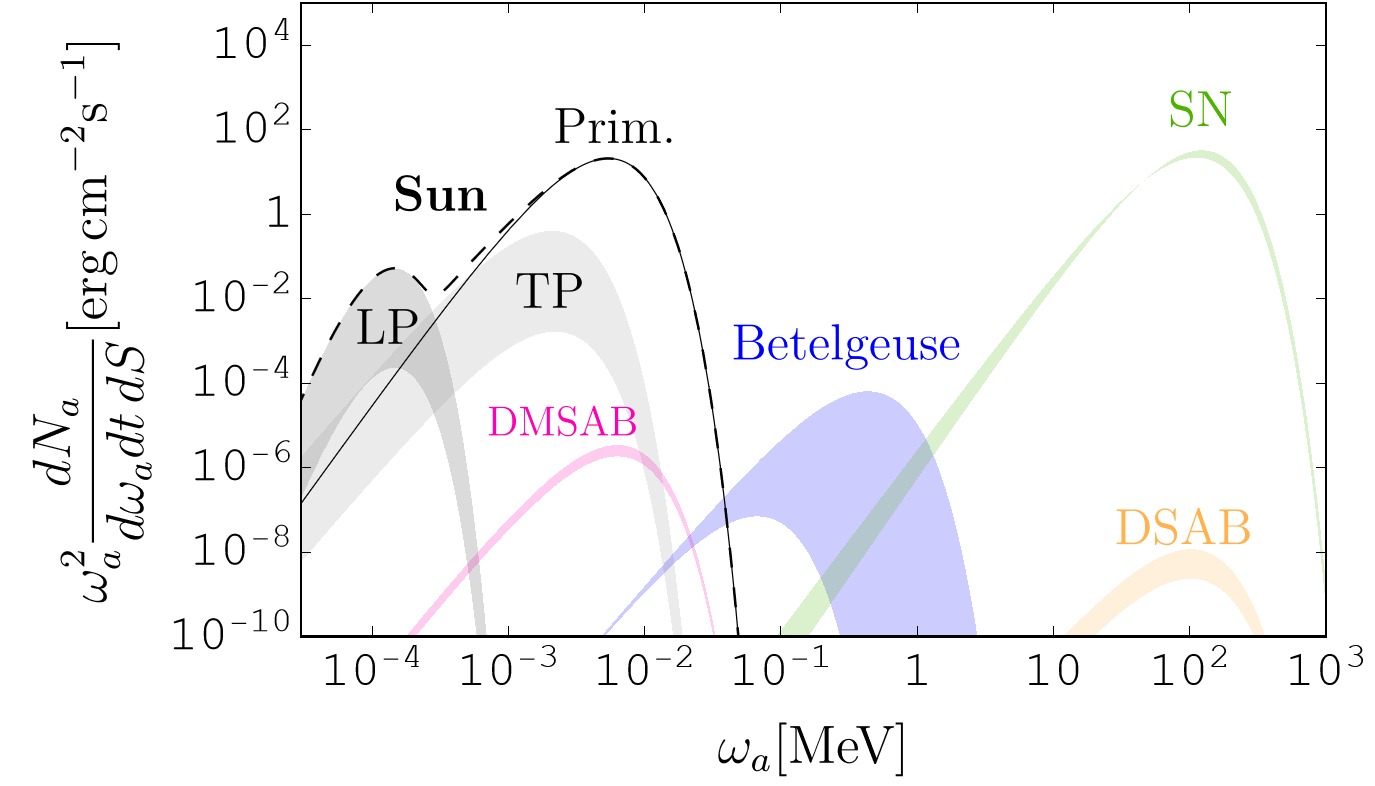}
    \caption{Axion energy flux at Earth as a function of the axion energy $\omega_a$, when considering axions coupled to photons and produced by different sources. Here, an axion-photon coupling $g_{a\gamma}=10^{-11}$~GeV$^{-1}$ and an axion mass $m_a \ll {\rm eV}$ are assumed. See Ref.~\cite{Carenza:2024ehj} for more details. Figure adapted from Ref.~\cite{Carenza:2024ehj}.}
    \label{fig:axion_flux}
  \end{center}
\end{figure}

Astrophysical systems provide exceptional laboratories for probing WISPs, offering unique insights beyond those accessible through terrestrial experiments. Indeed, WISPs can be copiously produced in stars. As an example, Fig.~\ref{fig:axion_flux} shows the expected axion energy flux from different stellar
systems, assuming an axion-photon interaction smaller than current experimental constraints. The presence of WISPs in stellar interiors can alter standard stellar evolution and lead to detectable signals on Earth. Observations of compact stars, supernovae, and other astrophysical environments have placed some of the strongest constraints on WISP properties, particularly for axions and ALPs~\cite{Caputo:2024oqc,Carenza:2024ehj}.

Among the most notable constraints are those derived from red giants, horizontal branch stars, and white dwarfs, which have set strong limits on axion-photon and axion-electron couplings. Supernova explosions, such as SN 1987A, have provided insights into WISPs up to masses of \(\mathcal{O}(100)\) MeV through analyses of the neutrino burst and associated gamma-ray emission. 
Additional constraints have emerged from neutron star cooling, stellar birefringence, superradiance, binary mergers and other astrophysical transients, each offering complementary probes to traditional laboratory experiments.

Indirect detection methods, including high-energy astrophysical observations of gamma rays, X-rays, and neutrinos, have further refined the parameter space for WISPs, including its viability as DM candidate. Observations of supernova neutrinos, in particular, offer an invaluable avenue for studying novel interactions of WISPs in extreme astrophysical conditions. Furthermore, the study of neutron stars and white dwarfs continues to provide a promising approach to identifying anomalous cooling behaviors that could hint at WISP emission.

Despite this progress, significant observational challenges remain. One of the most pressing issues is the so-called ``MeV sensitivity Gap'', a range of photon energies where current observational capabilities are severely limited. 
The development of next-generation space-based telescopes, such as COSI (Compton Spectrometer and Imager)~\cite{Caputo:2022dkz} and a future European mission akin to the ASTRO-MeV and e-ASTROGAM concepts, is essential
to bridging this gap and providing unprecedented insights into photon-WISP interactions. These instruments will enable detailed studies of WISP-induced gamma-ray signatures from supernovae, neutron stars, and other astrophysical sources.

Advancing our understanding of WISPs requires a coordinated effort in both theoretical modeling and instrumental development. 
Astrophysical missions such as JWST  (James Webb Space
Telescope)~\cite{Roy:2023omw} and the future NewAthena (Advanced Telescope for High-ENergy Astrophysics)~\cite{Sisk-Reynes:2022sqd,Cruise:2024mgo}, as well as ground-based observatories such as CTAO (Cherenkov Telescope Array Observatory)~\cite{CTA:2020hii}, will offer
complementary perspectives for WISP searches, providing crucial multi-wavelength observations. Additionally, gravitational wave observatories present an emerging avenue for probing the influence of ultra-light dark matter (ULDM) on wave propagation and generation~\cite{Duque:2023seg}.

Theoretical advancements will also play a key role in interpreting observational data and refining WISP constraints. Improved models of axion production and propagation in astrophysical magnetic fields are necessary to understand conversion mechanisms occurring in stellar atmospheres, as well as the effects of strong magnetic fields near magnetars. Furthermore, novel approaches are required to establish more direct connections between WISP phenomenology and fundamental parameters, such as axion-photon and axion-electron couplings, particularly in extreme astrophysical environments where these interactions are expected to be enhanced. These refinements will be critical for bridging theoretical predictions with observational efforts, strengthening the overall framework for WISP searches.

To fully exploit the potential of astrophysical WISP searches, it is crucial to prioritize the development of space-based MeV-range detectors, strengthen collaborations between astrophysics and particle physics communities, and emphasize the unique role of astrophysical systems in extending the search for new physics beyond the Standard Model. Astrophysics remains at the forefront of unveiling the nature of WISPs, reinforcing its essential complementarity to laboratory efforts.

\section{WISPS DIRECT SEARCHES}
\label{sec:WP4}


The last 15 years have seen an increasing interest in the direct search for WISPs, through broad range of new experiments designed to detect their non-zero coupling to SM particles and can be classified in:
\begin{description}
    \item[Haloscopes] Experiments designed to detect the WISP DM halo in our galaxy.
    \item[Helioscopes] Experiments designed to detect WISPs produced in the sun.
    \item[Pure lab experiments] Experiments designed to detect effects induced by WISP generated in the lab.
    \item[Fixed target and beam dump experiment]  Experiments designed to detect WISPs generated in the collision of an accelerated beam on a target.
\end{description}
Beside these, there are non-WISP-focused experiments with ability to detect WISPs. These are either collider experiments, experiments designed to detect WIMPs, gravitational wave interferometers, or high precision experiments able to detect small deviations from theoretically well predictable observables. 
\begin{figure}[!t]
  \begin{center}
    \includegraphics[totalheight=7cm]{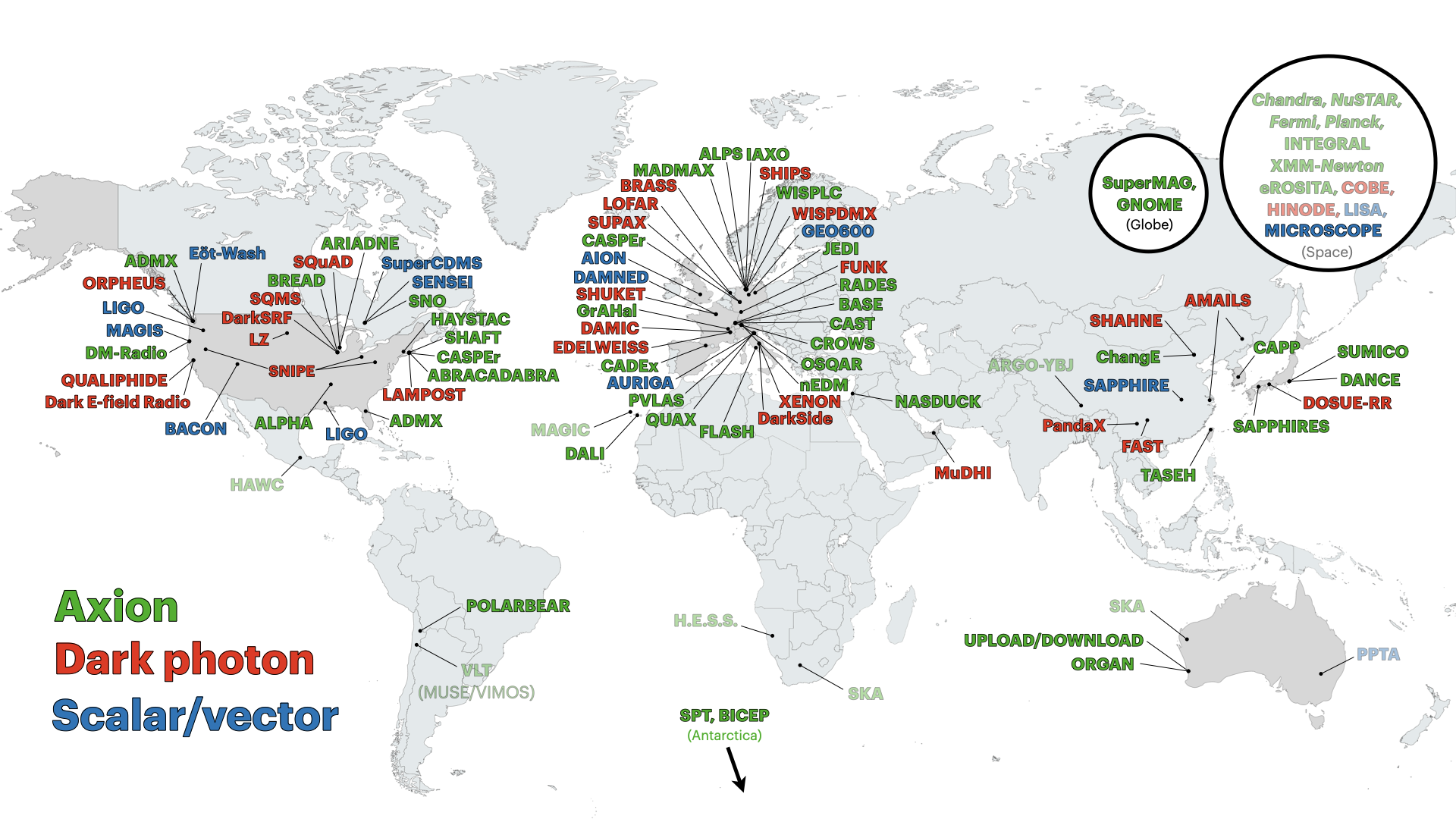}
    \caption{ Map of existing and planned WISPs experiments (\href{https://cajohare.github.io/AxionLimits/}{AxionLimits}). An interacting version is available at this \href{https://ggrillidc.github.io/experiments-map/}{link}. }
    \label{fig:AxionMapp}
  \end{center}
\end{figure}

\begin{figure}[htbp]
  \begin{center}
    \includegraphics[totalheight=8 cm]{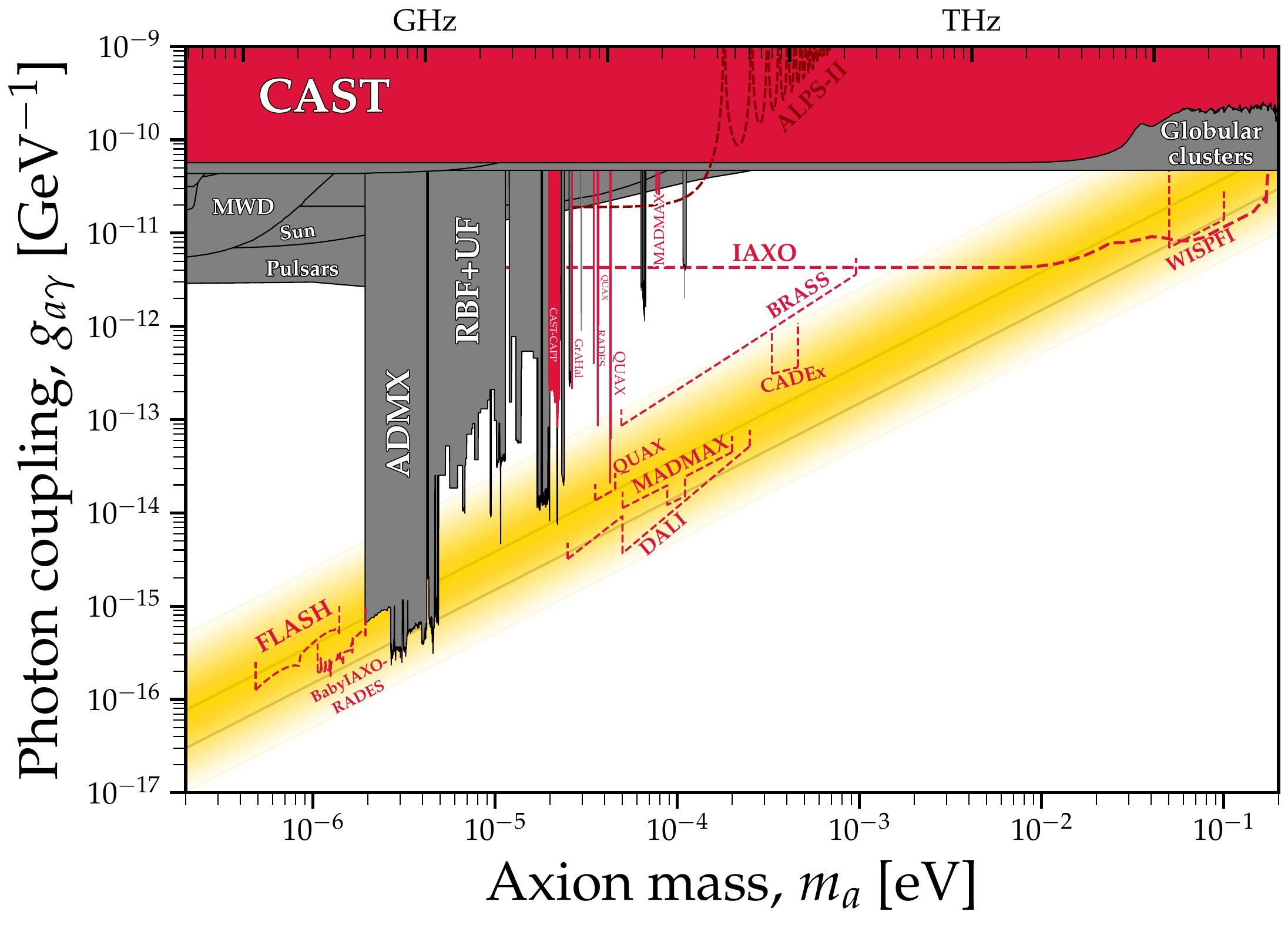}
    \caption{Projected limits on $g_{a\gamma}$ coupling as a function of the axion mass $m_a$ showing the present (plain red) and planned (dashed red lines) contribution from European experiments. The yellow band shows the region predicted by QCD-axion models.}
    \label{fig:AxionLimits-agg-proj}
  \end{center}
\end{figure}

The liveliness of this sector is particularly evident in Europe where there is a large and diverse WISPs research program with great impact and discovery potential. Dozens of experiments are, in fact, underway or being set up in universities and national laboratories, as highlighted in Fig.~\ref{fig:AxionMapp}.   This more than decade-long program of complementary research will probe much of the parameter space for various light DM models. As an example, the future impact of European research on the search for the QCD axion, one of the most theoretically motivated models, is exemplified in Fig.~\ref{fig:AxionLimits-agg-proj}. 
Many experiments share common technologies and would benefit from a more coordinated and structured approach.
European leadership in this field is built upon decades of expertise in
detector development, cryogenics, superconducting magnets, and radiofrequency technologies. The strength of the European effort lies in its diversity of experimental approaches, allowing for complementary
coverage of the parameter space while providing critical crossvalidation of potential signals.
Recent technological breakthroughs have enabled experiments to push into previously unexplored regions. For instance, advances in superconducting quantum-limited amplifiers have dramatically improved the sensitivity of haloscopes experiments like QUAX, while developments in high-field magnets have expanded the reach of helioscopes experiments
such as IAXO, the successor to well-known CAST. The MADMAX experiment represents an innovative approach using
dielectric haloscope to probe the axion mass range around 100 $\mu$eV, a
particularly challenging region. Meanwhile, BRASS and DALI are pioneering new techniques for probing the higher mass range,
complementing efforts like FLASH which targets the lower frequencies. Coordination between these diverse experimental efforts would yield significant benefits. Shared expertise in cryogenics, magnet technology, and low-noise electronics would accelerate development cycles. Common data analysis frameworks would strengthen the robustness of results. Furthermore, a unified approach to theoretical interpretation would provide clearer guidance for future experimental directions.

\section{Summary}

The last decade witnessed a growing interest in WISPs searches. A vivid research activity flourished around this topic ranging from WISP model building to impact in cosmological and astrophysical observations. 
An
increasing number of experiments have been proposed employing the most diverse techniques, with the goal to probe large regions of the WISP parameter space for different models.
WISP researchers constitute a relatively young community, where an important effort has been dedicated to research coordination and capacity building. In this sense, the COST Action \emph{Cosmic WISPers} aims to strongly stimulate this fascinating field of research which promotes WISP studies in a synergistic way to maximize the impact of the results and outline
a roadmap for WISP discovery in Europe.
In order to optimize the potential of these activities,
the proposed roadmap needs support by integrating WISP searches in the current update process of ESPP. Such a step is a prerequisite to further strive for the challenging realizations
of large-scale flagship projects like IAXO and MADMAX.
These efforts together with a worldwide variety  of smaller-scale experiments will provide a strong opportunity
for one or more game-changing discoveries, based on the WISP paradigm that may provide an answer to many 
unsolved puzzles of the Dark Universe.

\acknowledgements

This article is based upon work from COST Action COSMIC WISPers CA21106, supported by COST (European Cooperation in Science and Technology).

\bibliographystyle{bibi_new.bst}
\bibliography{reference_new.bib}

\section*{Endorsers}
\label{sec:endorse}
\addcontentsline{toc}{section}{\nameref{sec:endorse}}
\begin{longtable}{l l}
FIRST AND LAST NAME &     $ $ $ $ $ $ INSTITUTE \\
$ $ & $ $ \\
Aldo Serenelli &  $ $ $ $ $ $ Institute of Space Sciences (ICE, CSIC) \\
Alejandro Diaz-Morcillo &  $ $ $ $ $ $ Universidad Politecnica de Cartagena \\
Alessandro Melchiorri &  $ $ $ $ $ $ University of Rome, Sapienza, Italy \\
Ali Övgün &  $ $ $ $ $ $ Eastern Mediterranean University \\
Amedeo Maria Favitta &  $ $  $ $ $ $ $ $University of Palermo, Italy, INFN Catania, Italy \\
Andreas Ringwald & $ $  $ $ $ $ Deutsches Elektronen-Synchrotron DESY, Hamburg, Germany \\
Andreas Schachner &  $ $ $ $ $ $ Ludwig-Maximilian-University Munich\\
Ángeles Pérez García  &  $ $ $ $ $ $ University of Salamanca, Spain\\
Axel Lindner &  $ $ $ $ $ $ Deutsches Elektronen-Synchrotron DESY,  Hamburg, Germany \\
Benito Gimeno &  $ $ $ $ $ $ Instituto de Fisica Corpuscular IFIC (CSIC - University of Valencia), Spain \\
Beyhan Puliçe &  $ $ $ $ $ $ Sabancı University, Türkiye; The Open University of Israel\\
Ciaran A. J. O’Hare &  $ $ $ $ $ $  University of Sydney, Australia \\
Conrado Albertus Torres &  $ $ $ $ $ $ University of Salamanca, Spain \\
Cristina Margalejo &   $ $ $ $ $ $ Universidad de Zaragoza, Spain \\
Daniel Gavilan-Martin &  $ $ $ $ $ $ Helmholtz Institute and Johannes Gutenberg-Universität Mainz, Germany\\
Dmitry Budker &  $ $ $ $ $ $ Helmholtz Institute and JGU Mainz, Germany and University of California at Berkeley \\
Emmanuel N. Saridakis &  $ $ $ $ $ $ $ $National Observatory of Athens, Greece\\
Enrico Nardi &  $ $ $ $ $ $ NICPB, Tallinn, Estonia \\
Federico Mescia &  $ $ $ $ $ $ INFN LNF - University of Barcelona \\
Federico R.~Urban &  $ $  $ $ $ $ CEICO,  Czech Academy of Sciences, Prague \\
Francisco Rodríguez Candón &  $ $ $ $ $ $ Universidad de Zaragoza, Spain and TU Dortmund, Dortmund, Germany \\
Francesca Chadha-Day &  $ $ $ $ $ $ Durham University \\
Francesca Di Nella &  $ $ $ $ $ $ University of Rome Sapienza, Italy\\
Gaetano Lambiase &  $ $ $ $ $ $ Dipartimento di Fisica E.R. Caianiello, Universit{\`a} degli Studi di Salerno, Italy\\
Georg Raffelt & $ $ $ $ $ $ Max Planck Institute for Physics, Garching, Germany\\
Giuseppe Gaetano Luciano & $ $ $ $ $ $ $ $Escola Politecnica Superior, Universidad de Lleida\\
Giuseppe Messineo &  $ $ $ $ $ $ INFN Padova, Italy \\
G\"unter Sigl &  $ $ $ $ $ $ University of Hamburg, Germany\\
Hannah Banks &  $ $ $ $ $ $ DAMTP, University of Cambridge, United Kingdom \\
Hugo Ter\c{c}as   &  $ $ $ $ $ $ Instituto Superior de Engenharia de Lisboa \\
Igor G. Irastorza &  $ $ $ $ $ $  Universidad de Zaragoza, Spain \\
Ioannis D. Gialamas &  $ $ $ $ $ $ NICPB, Tallinn, Estonia \\
\.{I}zzet Sakall{\i} &  $ $ $ $ $ $ Eastern Mediterranean University, Turkey \\
Jaime Ruz &  $ $ $ $ $ $  TU Dortmund, Dortmund, Germany \\
Jordi Miralda-Escudé &  $ $ $ $ $ $ ICREA, ICCUB - University of Barcelona \\
Jose Cembranos &  $ $ $ $ $ $ IPARCOS-Universidad Complutense de Madrid \\
Julia K. Vogel &  $ $ $ $ $ $  TU Dortmund, Dortmund, Germany \\
Luca Brunelli &  $ $ $ $ $ $ University of Bologna \\
Luca Di Luzio &  $ $ $ $ $ $ INFN Padova \\ 
Luca Visinelli &  $ $ $ $ $ $ Universit{\`a} degli Studi di Salerno\\
Mario Ramos-Hamud &  $ $ $ $ $ $ DAMTP-University of Cambridge\\
Marios Maroudas &  $ $ $ $ $ $ Institute for Experimental Physics, University of Hamburg, Hamburg, Germany \\
Marta Fuentes Zamoro & $ $  $ $ $ $ $ $Universidad Aut\'onoma de Madrid and Instituto de F\'isica Te\'orica UAM/CSIC, Spain\\
Michele Gallinaro &  $ $ $ $ $ $ $ $ LIP Lisbon, Portugal \\
Miha Nemevšek &  $ $ $ $ $ $  University of Ljubljana and Institut Jožef Stefan, Ljubljana, Slovenia \\
Mimoza Hafizi  &  $ $ $ $ $ $  University of Tirana, Albania \\
Olga Mena  &  $ $ $ $ $ $ Instituto de F\'isica Corpuscular (IFIC), CSIC -- Universitat de Val\`encia \\ 
Olympia Maliaka &  $ $ $ $ $ $ Helmholtz Institute and JGU Mainz, Germany \\
Patricia Diego-Palazuelos & $ $   $ $ $ $ $ $Max Planck Institute for Astrophysics, Garching, Germany  \\
Pierluca Carenza & $ $  $ $ $ $ Stockholm University, Sweden\\
Rafid H. Dejrah &  $ $ $ $ $ $ Department of Physics, Faculty of Sciences, Ankara University, 06100, Ankara, T{u}rkiye\\
Reggie C. Pantig &  $ $ $ $ $ $ Mapúa University \\
Štěpán Kunc &  $ $ $ $ $ $ TUL Liberec, Czech \\
Thomas Schwetz &  $ $ $ $ $ $ Karlsruhe Institute of Technology \\
Thong T.Q. Nguyen & $ $  $ $ $ $ Stockholm University and The Oskar Klein Centre for Cosmoparticle Physics, Alba Nova, Stockholm, Sweden \\
Venelin Kozhuharov &  $ $ $ $ $ $ Faculty of Physics, Sofia University, Bulgaria \\
Venus Keus &  $ $ $ $ $ $ Dublin Institute for Advanced Studies, Dublin, Ireland\\
Victor Ovidiu Slupic & $ $ $ $ $ $ $ $University of Bucharest, Romania, IFIN-HH, Romania \\
Vasiliki A.\ Mitsou &  $ $ $ $ $ $ Instituto de F\'isica Corpuscular (IFIC), CSIC -- Universitat de Val\`encia 
\end{longtable}

\end{document}